\documentclass[cits]{PoS}

\usepackage{amssymb, fontenc, times, mathptmx, graphicx}
\usepackage{amsmath,mathrsfs, braket, caption, subcaption}

\DeclareMathOperator{\Tr}{Tr}

\newcommand*{\diff}{\mathop{}\!\mathrm{d}}

\newcommand*\fundiff[1]{\mathop{}\!\mathrm{D}#1\,}

\title{Lattice QCD$_2$ effective action with Bogoliubov transformations}

\ShortTitle{Lattice QCD$_2$ effective action with Bogoliubov transformations}

\author{Sergio Caracciolo\\
        Dipartimento di Fisica, Universit\`a degli Studi di Milano,\\
        and INFN, Via Celoria 16, I-20133 Milan, Italy\\
        E-mail: \email{sergio.caracciolo@mi.infn.it}}

\author{\speaker{Mauro Pastore}\\
        Dipartimento di Fisica, Universit\`a degli Studi di Milano,\\
        and INFN, Via Celoria 16, I-20133 Milan, Italy\\
        E-mail: \email{mauro.pastore@unimi.it}}

\abstract{In the Wilson's lattice formulation of QCD, a fermionic Fock space of states can be explicitly built at each time slice using canonical creation and annihilation operators. The partition function $Z$ is then represented as the trace of the transfer matrix, and its usual functional representation as a path integral of $\exp(- S)$ can be recovered in a standard way. However, applying a Bogoliubov transformation on the canonical operators before passing to the functional formalism, we can isolate a vacuum contribution in the resulting action which depends only on the parameters of the transformation and fixes them via a variational principle. Then, inserting in the trace defining $Z$ an operator projecting on the mesons subspace at each time slice and making the physical assumption that the true partition function is well approximate by the projected one, we can also write an effective quadratic action for mesons. We tested the method in the renowned 't~Hooft model, namely QCD in two spacetime dimensions for large number of colours, in Coulomb gauge.}

\FullConference{XIII Quark Confinement and the Hadron Spectrum - Confinement2018\\
		31 July - 6 August 2018\\
		Maynooth University, Ireland}

\begin{document}

\section{Motivation}
Confinement implies that the relevant degrees of freedom in QCD at low energy are not the fundamental ones (quarks and gluons), but effective composites, as the mesons. The situation shares some affinity with the BCS theory of superconductivity, where the macroscopic behaviour of the model cannot be explained in terms of non-interactive conduction electrons. In that framework, the superconductive phase is caused by the existence of an energy gap between the naive ground state and a more favorable one where the electrons are coupled in \emph{Cooper pairs}. The first vacuum state is related to the other via a Bogoliubov transformation, which maps the set of canonical operators of the fermions into the one of the quasiparticles, through a non trivial mixing between creators and annihilators.

Except for some remarkable cases that we do not have space to mention here (see \cite{reinhardt:2018dhg}), the application of this approach to QCD confinement, in spite of its intuitiveness, has been somewhat limited, because of the difficulty to work with canonical formalism in the relativistic setting of gauge theories.\footnote{Of course, Bogoliubov's ideas have been applied extensively in QCD to study colour superconductivity, but that's a different problem.} Some years ago, in~\cite{caracciolo:2006wc,palumbo:2007ax,caracciolo:2008ag,caracciolo:2010rm,caracciolo:2010xe}, a method has been developed to pass from a lattice operatorial description, where the Bogoliubov transformations can be implemented, to a functional formalism, using well-known tools as the expansion of the transfer matrix on a suitable basis of coherent states. In this way a theory of quasiparticle excitations above a non perturbative vacuum can be formulated. Moreover, the theory can be completely bosonized through a projection over the subspace of the relevant composite degrees of freedom, so that the description in term of effective fields arise from the fundamental fermionic theory in a coherent fashion.

Here we apply this method to the \emph{'t~Hooft model}~\cite{tHooft:1974pnl}, that is QCD$_2$ for large number of colours, where canonical formalism has been already applied fruitfully in the past, for example in~\cite{bardeen:1988mh,kalashnikova:2001df}, and where we can carry out our program to the end. In section \ref{sec:qcd2} we present the model, in Wilson's lattice formulation, in Coulomb gauge and in transfer matrix formalism. In section \ref{sec:effective} we find the gap equation in our framework and build a theory of quasifermions. In section \ref{sec:mesons} we show how an effective theory for mesons can be formulated in this setting. For details, we refer to~\cite{cp2018}.

\section{QCD$_{2}$ in Wilson's lattice formulation}
\label{sec:qcd2}

\subsection{Wilson action, Coulomb gauge}
\label{subsec:QCD2}
In functional formalism, the model is described by the Euclidean action $
S = S_F + S_G \, .
$
On a 2D rectangular lattice with spacings $a_0$ and $a_1$, the fermion part of the action is
\begin{multline}
S_F = a_0 a_1 \sum_{x\in (a_0\mathbb{Z})\times(a_1\mathbb{Z})}\Biggl\{\left(m+\frac{r_0}{a_0}+\frac{r_1}{a_1}\right)\bar{\psi}(x)\psi(x) \\
-\sum_{\mu=0}^1\biggl[\bar{\psi}(x)\frac{r_\mu-\gamma_\mu}{2a_\mu}U_\mu (x)\psi(x+a_\mu\hat{\mu}) 
+ \bar{\psi}(x+a_\mu\hat{\mu})\frac{r_\mu+\gamma_\mu}{2a_\mu}U^\dagger_\mu (x) \psi(x)\biggr]
\Biggr\}\, .
\label{eq:QCD2_action_fermions}
\end{multline}
The terms proportional to the Wilson parameters $r_\mu$ are introduced to solve the fermion doubling problem. The fermion fields are in the fundamental representation of the gauge group SU($N_c$), and the corresponding link variables are 
$
U_\mu(x) = \exp{\left[iga_\mu A_\mu(x) \right]}
$, with $A_\mu$ algebra-valued Hermitian fields and $g$ coupling constant. The pure gauge action is
\begin{equation}
S_G = \frac{1}{a_0a_1}\sum_P\frac{1}{g^2}\left[2N_c - \Tr\left(U_P + U_P^\dagger\right)\right] \, ,
\label{eq:QCD2_action_gauge}
\end{equation}
where, as usual, $P$ denotes an elementary plaquette and $U_P$ the ordered product of link variables around it.

Since~\cite{tHooft:1974pnl}, we know that the model can be solved in the \emph{'t~Hooft limit}, which is the double-scaling (large-$N_c$, weak -coupling) limit
$N_c \to \infty$, $\,g\to 0$, $\,g^2 N_c $ fixed. We will work, as in~\cite{bars:1977ud, bardeen:1988mh, kalashnikova:2001df}, in Coulomb gauge, which in 2 spacetime dimensions takes the simple form
\begin{equation}
U_1(x) = 1 \quad\iff\quad A_1(x) = 0 \,.
\label{eq:QCD2_Coulomb_gauge}
\end{equation}
In the weak-coupling approximation, the remaining link variable can be expanded as
\begin{equation}
U_0 = 1 + iga_0A_0 - \frac{1}{2} g^2 a_0^2 A_0^2 + \cdots \,,
\label{eq:QCD2_U0_weak}
\end{equation}
so, from \eqref{eq:QCD2_action_gauge}, we obtain the free gauge propagator $G^{ab}_{00}(x,y) = \braket{A_0^a(x) A^b_0(y)}$, where
\begin{equation}
G^{ab}_{00}(x,y) = \delta_{ab}\frac{\delta_{x^0 , y^0}}{a_0}\int_{-\pi/a_1}^{\pi/a_1} \frac{\diff p}{2\pi}\,\frac{e^{ip\left(x^1-y^1\right)}}{\hat{p}^2} \quad\xrightarrow{\substack{a_0\to 0\\a_1\to 0}} \quad -\frac{1}{2}\delta_{ab}\delta(x^0 - y^0)\left|x^1 - y^1 \right| \,.
\label{eq:QCD2_gauge_propagator}
\end{equation}
Thus, in QCD$_2$ the interaction mediated by the gluons is instantaneous and confining at perturbative level already, because of the linearity of the Coulomb potential in the spatial separation.

\subsection{Canonical formalism and transfer matrix}
\label{subsec:transfer}
The quantum theory descends from the partition function
\begin{equation}
Z = \int \!\fundiff U\! \fundiff \psi\! \fundiff{\bar{\psi}}\, e^{-S[\psi,\bar{\psi},U]} 
=   \int \!\fundiff U\! e^{-S_G[U]} \, \int \!\fundiff \psi\! \fundiff{\bar{\psi}}\,  e^{-S_F[\psi,\bar{\psi},U]}
=  \int \!\fundiff U\! e^{-S_G[U]} \, Z_F \, ,
\label{eq:transfer_partition_functional}
\end{equation}
in path-integral representation. However, the fermionic Fock space of states can be constructed explicitly on the lattice. At a fixed time-slice, the ladder operators for fermions are, in the standard basis,
\begin{equation} 
\hat{\psi} =  \left(\begin{array}{c}\hat{u}  \\ \hat{v}^\dagger \end{array}\right)\, .
\label{eq:transfer_std_basis}
\end{equation}
The operator $\hat{u}^{\dagger\,i}_{p}$ ($\hat{v}^{\dagger\, i}_{p}$) creates a quark (antiquark) with spatial momentum $p$ and colour $i$. Given that, the space of states  is built acting with these ladder operators on the vacuum $\ket{0}$, such that
\begin{equation}
\hat{u} \ket{0} = 0 \,, \qquad \hat{v}\ket{0} = 0 \,.
\label{eq:transfer_vacuum}
\end{equation}
In this setting, the partition function can be expressed as the trace of the Boltzmann factor:
\begin{equation}
Z_F = \lim_{\beta \to \infty}\Tr^F\, e^{- \beta \hat{H}_F} = \Tr^F\, \prod_t J_t\hat{\mathcal{T}}_{t,t+1} \, .
\label{eq:transfer_partition_statistical_T}
\end{equation}
The operator $\hat{\mathcal{T}}_{t,t+1}$ is the \emph{transfer matrix}, which maps the Hilbert space defined at a certain time $t$ into the next one. In QCD, this operator is self-adjoint and strictly positive, as it has been proven long ago in~\cite{luescher:1976ms}, so a lattice Hamiltonian can be defined as in~\eqref{eq:transfer_partition_statistical_T}. Moreover, the transfer matrix takes the form
\begin{equation}
\hat{\mathcal{T}}_{t,t+1} =  \hat{T}^\dagger_t \hat{V}_t \hat{T}_{t+1} \,,
\label{eq:transfer_matrix}
\end{equation}
where
\begin{equation}
\hat{T}_{t} = \exp \left[-\hat{u}^\dagger M_t \hat{u} +  \hat{v} M_t \hat{v}^\dagger \right] \exp\left[\hat{v} N_t \hat{u} \right]\,,\qquad
\hat{V}_{t} = \exp \left[\hat{u}^\dagger \ln U_{0,t} \hat{u} -   \hat{v}\ln U^\dagger_{0,t}\hat{v}^\dagger \right]
\end{equation}
and $M_t$, $N_t$, $U_{0,t}$ are matrices in internal (colour) and space (but not time, which is just a label) indices, with the last one defined as 
$
[U_{\mu,t}]_{n_1 n_2}\equiv  \delta_{n_1 n_2} (U_\mu)_{n_1,t}
$. The extra factor $J_t$ appearing in~(\ref{eq:transfer_partition_statistical_T}) is simply
\begin{equation}
J_t = \exp{\left[\Tr \left(M_t + M^\dagger_t \right)\right]} \, .
\end{equation}

The bridge between the functional representation~\eqref{eq:transfer_partition_functional} and the operatorial one of the partition function goes through the choice of a basis of the Hilbert space on which the trace in~\eqref{eq:transfer_partition_statistical_T} can be evaluated explicitly. In particular, it is convenient to take the basis of \emph{canonical coherent states}, defined by
\begin{equation}
\ket{\rho \sigma} = \exp\left(-\rho \hat{u}^\dagger - \sigma \hat{v}^\dagger\right)\ket{0} \,,
\label{eq:transfer_coherent}
\end{equation}
where $\rho$, $\sigma$ are anticommuting (Grassmannian) symbols, such that\footnote{Note that, because of the fact that they are associated to Grassmannian eigenvalues, the fermionic coherent states are not, strictly speaking, part of the Hilbert space. However, all can be defined in a proper way, using the holomorphic formalism.}
\begin{equation}
\hat{u} \ket{\rho\sigma} =  \rho \ket{\rho\sigma}, \qquad  \hat{v} \ket{\rho\sigma}  =  \sigma  \ket{\rho\sigma}\,.
\end{equation}
Inserting the identity operator
\begin{equation}
\hat{\mathbb{I}}= \int \prod_K \diff \rho^\dagger_K  \diff \rho_K \diff \sigma^\dagger_K  \diff \sigma_K \,e^{- \rho^\dagger \rho-\sigma^\dagger \sigma} \ket{\rho\sigma}\!\bra{\rho \sigma}
\label{eq:transfer_identity}
\end{equation}
in the trace at each time slice, grouping the Grassmann fields in a Dirac spinor
\begin{equation}
\psi_{t,p}^i = \begin{pmatrix}
\rho_{t,p}^i\\
{\sigma^{\dagger i}_{t,p}}
\end{pmatrix}
\end{equation}
and confronting with the functional integral representation~\eqref{eq:transfer_partition_functional}, the form of the matrices
\begin{subequations}
\begin{align}
M_t &= \frac{1}{2}\log \frac{B_t}{2\kappa_0} \equiv  \frac{1}{2}\log \frac{\mathbb{I} - \kappa_1 r_1 \left(U_{1,t} T_1 + T^\dagger_1 U^\dagger_{1,t} \right)}{2\kappa_0} \, , \label{eq:transfer_M}\\
N_t &= -i\kappa_1 B_t^{-1/2} \left(U_{1,t} T_1^{(+)} - T_1^{(-)} U^\dagger_{1,t}\right)  B_t^{-1/2} \label{eq:transfer_N}
\end{align}
\label{eq:transfer_MN}%
\end{subequations}
follows, with $\kappa_0$, $\kappa_1$ hopping parameters and $T^{(\pm)}$ shift operators in the spatial direction. In weak-coupling and at lowest order in the lattice spacings, we can write, in momentum space,
\begin{subequations}
\begin{align}
e^{M}(q) &= 1 + \frac{ma_0}{2} + O(a_0 a_1)\, ,\\
N(q) &= a_0 q + O(a_0 a_1)\, ,
\end{align}
\label{eq:transfer_MN_expansion}%
\end{subequations}
because of translational invariance.

\section{Effective action with Bogoliubov transformations}
\label{sec:effective}

An equivalent description of the model can be obtained after a unitary linear transformation on the vector space of the Dirac operators. This is realized defining a new set of canonical operators at each time slice as
\begin{equation}
\begin{aligned}
&\hat{a} = R^{1/2}\left(\hat{u} - F^\dagger \hat{v}^\dagger\right) \,,\\
&\hat{a}^\dagger = \left(\hat{u}^\dagger - \hat{v}F \right)R^{1/2} \,,
\end{aligned}
\qquad
\begin{aligned}
&\hat{b} = \left(\hat{v} +  \hat{u}^\dagger F^\dagger\right) \mathring{R}^{1/2}\,, \\
&\hat{b}^\dagger = \mathring{R}^{1/2}\left(\hat{v}^\dagger + F \hat{u}\right)\,,
\end{aligned}
\label{eq:effective_bogoliubov}
\end{equation}
where $F$ is an arbitrary matrix in space and colour indices and
\begin{equation}
R = \left(1+F^\dagger F \right)^{-1}\,, \qquad
\mathring{R} = \left(1+ F F^\dagger \right)^{-1}\,,
\label{eq:effective_R}
\end{equation}
assuring the new operators, called \emph{quasiparticle operators}, still to be canonical (see~\cite{caracciolo:2008ag} for details). In QCD$_2$, assuming the transformation to be translational invariant and proportional to the identity in colour space, the matrices $F$ can be parametrized with a single angular variable for each momentum:
\begin{equation}
F(p,q) = 2\pi a_1 \delta(p+q) F(q) \,,\qquad \text{with}\quad F(q) = \tan \frac{\theta_q}{2} = F^\dagger(q)\,.
\label{eq:effective_theta}%
\end{equation}
The so-called Bogoliubov transformation \eqref{eq:effective_bogoliubov} was first introduced to study superfluidity and superconductivity in condensed matter, where in the low-temperature phase the systems are in a ground states with energy strictly \emph{below} the one of the naive vacuum of the original fermionic excitations. Indeed, because of the non trivial mixing between creators and annihilators in the definition, the quasiparticle operators are associated to a new vacuum state in the Hilbert space, which is
\begin{equation}
\ket{F} = \exp \left( \hat{u}^\dagger F^\dagger \hat{v}^\dagger\right)\ket{0} \qquad \text{such that }\quad \hat{a} \ket{F} = 0\,, \quad \hat{b}  \ket{F} = 0 \,.
\label{eq:effective_vacuum}
\end{equation}
From this, a new basis of coherent states can be built as
\begin{equation}
\ket{\alpha\beta;F} = \exp\left(-\alpha \hat{a}^\dagger - \beta \hat{b}^\dagger \right) \ket{F}
= \exp\left(-\alpha \hat{a}^\dagger - \beta \hat{b}^\dagger \right) \exp \left( \hat{u}^\dagger F^\dagger\hat{v}^\dagger\right)\ket{0}\, ,
\label{eq:effective_coherent}
\end{equation}
such that, now,
\begin{equation}
\hat{a} \ket{\alpha\beta;F}  =  \alpha \ket{\alpha\beta;F}\,, \qquad  \hat{b} \ket{\alpha\beta;F}  =  \beta  \ket{\alpha\beta;F}\, .
\end{equation}
Operating as before, a new functional representation of the fermionic partition function is obtained in terms of the quasiparticle fields:
\begin{equation}
Z_F = \int \! \fundiff{\alpha^\dagger}\!\fundiff\alpha\! \fundiff{\beta^\dagger}\!\fundiff\beta\, e^{-S_0[U;F] - S_{QP}[\alpha,\beta,U;F]}\,.
\label{eq:effective_partition}
\end{equation}
In it we can recognize a term depending only on the parameters of the transformation, which is a sort of vacuum contribution to the action, and a fermionic action for the quasiparticle fields. Note that, at this level, no approximation has been introduced.

\subsection{Vacuum contribution}

\begin{figure}
\begin{minipage}[t]{0.49\textwidth}
\input{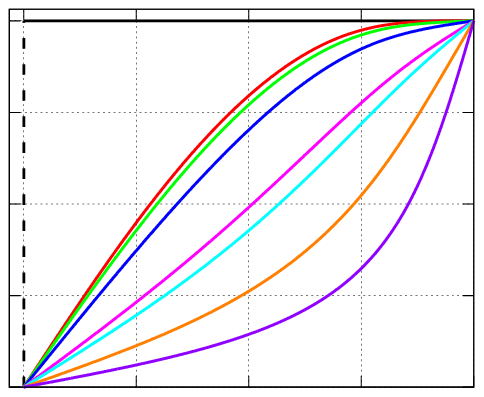}
\end{minipage}
\begin{minipage}[t]{0.49\textwidth}
\input{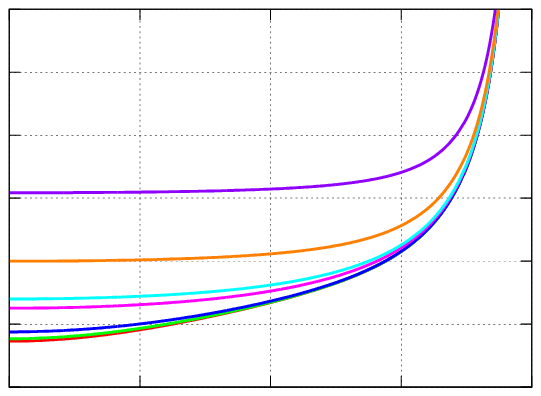}
\end{minipage}
\caption{Plot for the $\theta_p$ that solves the gap point equation (left) and the corresponding vacuum energy $\omega^0$ (right). On the left, the topmost, discontinuous line (in black) corresponds to the chiral limit of the free theory ($m=0$, $\gamma\to 0$); the following lower one (in red) corresponds to the chiral limit of the interacting theory ($m=0$, $\gamma=1$). The others, from the top to the bottom, correspond to different values of the mass in the interacting theory ($\gamma=1$) in the set $m\in \{0.045,0.18,0.749,1,2.11,4.23\}$, in unities of $\sqrt{2\gamma}$, to compare with references~\cite{li:1987hx,jia:2017uul}.}
\label{fig:gap}
\end{figure}

When the quasiparticle contribution is switched off, still the zero-point action $S_0$ is present in \eqref{eq:effective_partition}. A variational principle on it can be use to fix the parameters $F$ to the value $\bar{F}$ which implement the Bogoliubov transformation that produces the physical vacuum $\ket{\bar{F}}$, the state of mi\-ni\-mal energy. This request of extremality with respect of all possible vacua generated in this way translates into the gap equations
\begin{equation}
\left.\frac{\delta S_0}{\delta F_t}\right|_{F_t=\bar{F}_t} = 0 = \left.\frac{\delta S_0}{\delta F_t^\dagger}\right|_{F^\dagger_t=\bar{F}^\dagger_t} \,.
\label{eq:vacuum_gap}
\end{equation}
The dependence of $S_0$ on the gauge fields configuration, which changes in time, makes difficult to solve these equation in general, as noted in~\cite{caracciolo:2010rm}. However, because of the possibility to obtain non-trivial results in QCD$_2$ in the weak-coupling limit, a solution can be found \emph{on average}, that is after a perturbative expansion to second order in the coupling constant $g$ and an integration over the gauge fields. After the average, the vacuum action becomes
\begin{multline}
S_0[\theta] = \braket{S_0[\theta;A_0]} = - \frac{V N_c}{2\pi}\, \biggl[\int \!\diff q\left(m  \cos\theta_q +  q \sin\theta_q\right) \\
-\frac{ \alpha_s\left(N_c^2-1  \right)}{2 N_c}\int\!\diff q\int\!\diff k\,\frac{1}{(q-k)^2}\sin^2 \frac{\theta_q - \theta_k}{2}  \biggr] \,,
\label{eq:vacuum_action}
\end{multline}
where $V$ is the spacetime volume. Defining the vacuum dispersion relation
\begin{equation}
\omega^0_q[\theta] = m  \cos\theta_q +  q \sin\theta_q -\frac{\gamma}{2}\int\!\diff k\,\frac{1}{(q-k)^2}\sin^2 \frac{\theta_q - \theta_k}{2}
\label{eq:vacuum_energy}
\end{equation}
with
\begin{equation}
\gamma \equiv \alpha_s\frac{N_c^2-1  }{N_c} \underset{N_c\to \infty}{\longrightarrow} \alpha_s N_c \,,
\end{equation}
the equations \eqref{eq:vacuum_gap} become
\begin{equation}
\left.\frac{\diff}{\diff \theta_q} \omega_q^0[\theta]\right|_{\theta=\bar{\theta}} = -m  \sin\bar{\theta}_q +  q \cos\bar{\theta}_q -\frac{ \gamma}{2}\int\!\diff k\,\frac{\sin \left(\bar{\theta}_q - \bar{\theta}_k\right)}{(q-k)^2} = 0 \,.
\label{eq:vacuum_theta_saddle}
\end{equation}
These are the well-known gap equations for the 't~Hooft model, see~\cite{bars:1977ud}. The behaviour of the solutions and of the vacuum energy $\omega^0$ evaluated on them is plotted in Figure~\ref{fig:gap}. Note that, however, we are able to re-obtain them with a continuum limit of a more general lattice formulation.

\subsection{Quasiparticle action}

After the average over the gauge fields, the quasiparticle term in \eqref{eq:effective_partition} becomes a quartic action in the fermion fields. Evaluated on $\bar{\theta}$, it can be written as 
\begin{multline}
S_{QP}\left[\alpha,\beta;\bar{\theta}\right] 
= - a_0\sum_t \frac{1}{a_1}\int \frac{\diff q}{2\pi}
\Bigl[ \alpha^\dagger_t(q) \left(\partial^{(-)}_t - \omega^{QP}_q[\bar{\theta}]\right)\alpha_{t+1}(-q)\\
 - \beta_{t+1}(q)\left(-\partial_t^{(+)} -\omega^{QP}_q[\bar{\theta}]\right)\beta^\dagger_t(-q)\Bigr]
+ a_0 \sum_t \frac{1}{a^2_1} V_t\left[\alpha,\alpha^\dagger,\beta,\beta^\dagger;\bar{\theta}\right] \,,
\end{multline}
where the quasiparticle energies are
\begin{equation}
\omega^{QP}_q[\theta] = m \cos\theta_q +  q \sin\theta_q  + \frac{\gamma}{2}\int \frac{\diff k}{2\pi}\,\frac{\cos(\theta_q - \theta_k)}{(q-k)^2} 
\end{equation}
and $V_t$ is a quartic interaction potential produced by the gauge integration. Here confinement is understood as the infrared divergence of $\omega^{QP}_q[\bar{\theta}]$: in the low momentum phase the quasiparticles does not propagate individually because they would require an infinite energy to do so.

Note that the Bogoliubov transformation generates also the mixing quadratic terms
\begin{equation}
\frac{a_0}{a_1}\sum_t \int \frac{\diff q}{2\pi} \,I_q[\theta] \left[\beta_t(q)  \alpha_t(-q) + \alpha^\dagger_t(q) \beta^\dagger_t(-q)\right],
\end{equation}
with, after the average over gauge,
\begin{equation}
I_q[\theta] = - m \sin\theta_q + q \cos\theta_q 
- \gamma \int \!\diff k\,\frac{1}{(q-k)^2} \cos \frac{\theta_q - \theta_k}{2}\sin \frac{\theta_q - \theta_k}{2} \biggr] \,.
\end{equation}
However, as
$
I_q[\theta] = \diff \omega_q^0/\diff \theta_q
$, these terms are null in $\bar{\theta}$: in the theory built above the physical vacuum, quarks and antiquarks decouple at the quadratic level.

\section{Projecting on mesons}
\label{sec:mesons}
In the previous sections, we show how QCD$_2$ can be written in terms of quasiparticle excitations above the physical vacuum. However, the resulting theory is still a quartic fermionic theory, with all the complications that follow. In order to bosonize the model, we can impose an hypothesis of \emph{composite boson dominance}~\cite{palumbo:2005pq,palumbo:2007nn,caracciolo:2006wc}: the relevant states in the spectrum must be the quasiparticle condensates
\begin{equation}
\ket{\Phi; F} = \exp \bigl(\hat{a}^\dagger \Phi^\dagger \hat{b}^\dagger\bigr) \ket{F} \,,
\label{eq:mesons_condensate}
\end{equation}
with $\Phi$ structure matrices. This hypothesis is realized at the level of the partition function in canonical formalism \eqref{eq:transfer_partition_statistical_T}: the insertion of the operator
\begin{equation}
\hat{\mathcal{P}}[F] = \int  \frac{\left[\diff\Phi^\dagger \diff\Phi\right]}{\braket{\Phi;F|\Phi;F}} \ket{\Phi;F}\!\bra{\Phi;F} \,,
\label{eq:mesons_projector}
\end{equation}
which projects onto composites of the form \eqref{eq:mesons_condensate}, is required to not modify the fermionic partition function, that is
\begin{equation}
\begin{aligned}
Z_F\simeq Z_C &= \Tr^F \prod_t J_t\hat{\mathcal{P}}_t \hat{\mathcal{T}}_{t,t+1}\\
&=  \int \!\prod_t\left[\diff{\Phi^\dagger_t} \diff{\Phi_t}\right] J_t \frac{\braket{\Phi_t;F_t | \hat{\mathcal{T}}_{t,t+1} | \Phi_{t+1} ;F_{t+1}}}{\braket{\Phi_t;F_t |\Phi_t;F_t}}\\
&= \int \! \fundiff{\Phi^\dagger}\!\fundiff\Phi \, e^{-S_0[U;F] - S_M[\Phi,\Phi^\dagger,U;F]} \,.
\end{aligned}
\label{eq:mesons_partition}
\end{equation}
The vacuum contribution is unchanged, while the term $S_M$ is a complicated functional in the structure matrices. However, as the relevant composites for large $N_c$ are expected to be mesons, we can make the choice
\begin{equation}
\Phi = \mathbb{I}_{N_c} \frac{\phi}{\sqrt{N_c}} \,\,,
\end{equation}
so that the resulting condensates are colourless. For large $N_c$ and evaluated in $\bar{\theta}$, $S_M$ becomes the \emph{effective quadratic action for mesons}
\begin{equation}
\begin{aligned}
S_M[\phi,\phi^\dagger] = a_0 \sum_{t} \int \frac{\diff Q\diff q}{(2\pi)^2}
 &\Biggl\{\phi_t(Q-q,q)\left[ - \partial^{(-)}_0 + \left(\omega_{Q-q} + \omega_{q}\right)\right]\phi^\dagger_{t+1}(-q,-Q + q)\\
  - \frac{\gamma}{2}\int\frac{\diff q'}{\left(q-q'\right)^2}\biggl[ & 2\cos \frac{\theta_{q} -\theta_{q'} }{2} \cos \frac{\theta_{Q-q} - \theta_{Q-q'}}{2}
\phi_t(Q-q,q)\phi^\dagger_{t+1}(-q',-Q+q')\\
+&\sin \frac{\theta_{q} -\theta_{q'} }{2} \sin \frac{\theta_{Q-q} - \theta_{Q-q'}}{2} 
\phi_t(Q-q,q)\phi_{t}(-q',-Q+q')\\
+&\sin \frac{\theta_{q} -\theta_{q'} }{2} \sin \frac{\theta_{Q-q} - \theta_{Q-q'}}{2}
\phi^\dagger_t(Q-q,q)\phi^\dagger_{t}(-q',-Q+q') \biggr] \Biggr\}\,.
\end{aligned}
\label{eq:QCD2_action}
\end{equation}
We recognize this expression as the action, in holomorphic representation, arising from the effective Hamiltonian for the composites already found in \cite{bardeen:1988mh,kalashnikova:2001df} directly in the continuum. Separating the parts $\varphi^{(n)}(Q)$ depending only on the center-of-mass momenta, which will be identified with the physical mesons, from the one also depending on internal structure of the fields $\phi(p,q)$, this action can be diagonalized in a form
\begin{equation}
S_M = a_0 \sum_{t} \int \frac{\diff Q}{2\pi} \sum_n \varphi^{(n)}_t(Q) \left[ -\partial^{(-)}_0  + \lambda_Q^{(n)}\right]  {\varphi^{\dagger(n)}_{t+1}}(-Q)\,,
\label{eq:QCD2_structure_action_diagonal}
\end{equation}
requiring the structure functions to comply with the \emph{Bars-Green equations} for the mesonic spectrum, as in~\cite{bars:1977ud,kalashnikova:2001df}.

\section{Conclusions}
At the end of the day, starting from the fundamental theory of quarks and gluons in QCD$_2$, we obtained an effective theory for mesons on the lattice which reproduces, in the continuum limit, results well known from previous Hamiltonian canonical approach. In doing so, we earned a remarkable physical insight about the hypothesis of boson dominance, implemented through a projection on composite states.

In the future, 
we aim to study models at finite temperature and chemical potential, to verify, for example, how diquarks states can be realized using the approach in~\cite{caracciolo:2011aa}.

\bibliographystyle{JHEP}
\bibliography{biblio}

\end{document}